\documentstyle[12pt,psfig,epsf]{article}
\newcommand{\be}{\begin{equation}}
\newcommand{\ee}{\end{equation}}

\newcommand{\beqa}{\begin{eqnarray}}
\newcommand{\eeqa}{\end{eqnarray}}

\newcommand{\eqref}[1]{(\ref{#1})}

\def\boxit#1{\vbox{\hrule\hbox{\vrule\kern8pt
\vbox{\hbox{\kern8pt}\hbox{\vbox{#1}}\hbox{\kern8pt}}
\kern8pt\vrule}\hrule}}
\def\mathboxit#1{\vbox{\hrule\hbox{\vrule\kern8pt\vbox{\kern8pt
\hbox{$\displaystyle #1$}\kern8pt}\kern8pt\vrule}\hrule}}

\def\IB{\relax\hbox{$\inbar\kern-.3em{\rm B}$}}
\def\IC{\relax\hbox{$\inbar\kern-.3em{\rm C}$}}
\def\ID{\relax\hbox{$\inbar\kern-.3em{\rm D}$}}
\def\IE{\relax\hbox{$\inbar\kern-.3em{\rm E}$}}
\def\IF{\relax\hbox{$\inbar\kern-.3em{\rm F}$}}
\def\IG{\relax\hbox{$\inbar\kern-.3em{\rm G}$}}
\def\IGa{\relax\hbox{${\rm I}\kern-.18em\Gamma$}}
\def\IH{\relax{\rm I\kern-.18em H}}
\def\IK{\relax{\rm I\kern-.18em K}}
\def\IL{\relax{\rm I\kern-.18em L}}
\def\IP{\relax{\rm I\kern-.18em P}}
\def\IR{\relax{\rm I\kern-.18em R}}
\def\IZ{\relax\ifmmode\mathchoice
{\hbox{\cmss Z\kern-.4em Z}}{\hbox{\cmss Z\kern-.4em Z}}
{\lower.9pt\hbox{\cmsss Z\kern-.4em Z}} {\lower1.2pt\hbox{\cmsss
Z\kern-.4em Z}}\else{\cmss Z\kern-.4em Z}\fi}

\def\II{\relax{\rm I\kern-.18em I}}

\begin{document}

\hfill  NRCPS-HE-03-26

\vspace{5cm}
%\begin{titlepage}
%\title{
\begin{center}
{\large \bf Two-loop World-sheet Effective Action
}%title ends

\vspace{2cm}
%\author{

{\sl A.R.Fazio\footnote{email:~fazio@inp.demokritos.gr}
and G.K.Savvidy\footnote{email:~savvidy@inp.demokritos.gr}}\\
Demokritos National Research Center ,\\
Ag. Paraskevi, GR-15310 Athens, Hellenic Republic\\

%}%author ends
%}
%\date{}%in order NOT to write the date
%\maketitle
\end{center}
\vspace{60pt}

\centerline{{\bf Abstract}}

\vspace{12pt}

\noindent
We are studying quantum corrections in the earlier proposed
string theory based on world-sheet action which measures
the linear sizes of the surfaces. At classical level the string
tension is equal to zero and as it was demonstrated in the previous
studies one loop correction to the classical world-sheet action
generates Nambu-Goto area term, that is nonzero string tension.
We extend this analysis computing the world-sheet effective action in the
second order of the loop expansion.

%\end{abstract}
%\thispagestyle{empty}
%\end{titlepage}

\newpage

\pagestyle{plain}
%\pagenumbering{roman}
\section{Introduction}

   In \cite{geo,sav} authors suggested a model of string theory
based on world-sheet action which measures the linear
sizes of the surfaces, so-called gonihedric model. The world-sheet
action of the theory is defined by the integral:
\be\label{funaction}
S =m \cdot L= m \int_{\Sigma} d^{2}\zeta
\sqrt{g}\sqrt{K^{ia}_{a}K^{ib}_{b}},
\ee
where L measures the linear size of the
surface $\Sigma$, $m$ has dimension of mass and $K^{i}_{ab}$ is a second
fundamental form (extrinsic curvature)\footnote{It differs from the models
considered in the previous studies
\cite{polykov,kleinert,helfrich,weingarten}, because
the action has dimension of length $ L ~~ \propto ~~length $, it is
proportional to the linear size of the surface similar to the path
integral action.  This is in contrast with the previous proposals
where the extrinsic curvature term is a dimensionless functional
$S(extrinsic~curvature) \propto 1$.}.

At the classical level string tension  is equal
to zero $T_{classical}=0$, because the action (\ref{funaction}) is
equal to the perimeter of the flat Wilson loop $ S \rightarrow  m(R+T)$.
It was demonstrated in \cite{geo,ruben} that quantum fluctuations
generate dynamically the Nambu-Goto area term $A_{NG}$ in the world-sheet effective action
\be\label{qtension}
{{\cal S}}_{eff}  = S~~ + ~~T_{q} ~ A_{NG}~~ +~~ ...,
\ee
with non-zero string tension  $T_{q} =m~M$.
Dynamical generation of string tension $T_q$  has been found in the discrete
formulation of the theory when the action (\ref{funaction}) is
written for the triangulated surfaces \cite{geo}. In \cite{ruben}
the authors have computed one-loop effective lagrangian
${{\cal L}}_{eff}={{\cal L}}_0 + {{\cal L}}_1$ and found that
similar generation of non-zero string tension takes place in a
continuum formulation of the theory (the ${{\cal L}}_{eff}$
is a functional of the Lagrange multiplier $\lambda_{ab} = \lambda \eta_{ab}$).

Here we shall extend this analysis computing the world-sheet
effective Lagrangian in the second order of the loop
expansion~${{\cal L}}_{eff}={{\cal L}}_0 +
{{\cal L}}_1+ {{\cal L}}_2$.
We get the following result for the world-sheet effective
action up to two-loop order which depends on $\lambda$:
\beqa
{{\cal S}}_{eff} = m \left[ 1 + a ~{\lambda  \over m}
\left(\ln\left(\frac{\lambda}{M}\right)-1 \right) +
b~\left({\lambda \over m}\right)^2 ~
\ln^2 \left(\frac{\lambda}{M}\right) \right]
\int  d^{2}\zeta\sqrt{g}\sqrt{K^{ia}_{a}K^{ib}_{b}}
\eeqa
together with the term which is proportional to the Polyakov-Kleinert
action
\beqa
S_{2~PK} = {1\over  4\pi e^{2}_{eff}} \int d^{2}\zeta\sqrt{g}K^{ia}_{a}K^{ib}_{b}
,~~~~~~ {1\over e^{2}_{eff} }=a ~{\lambda \over m}   ~\ln^2
\left(\frac{\lambda}{M}\right) \nonumber
\eeqa
where $a=\frac{(D-3)}{4\pi},~~~b=- \frac{(D-3)(D-5)}{32\pi^2} $.
Quantum fluctuations generate high powers of
the extrinsic curvature tensor $K^{i}_{ab}$ in the form of
Polyakov-Kleinert action with the induced coupling constant $e^{2}_{eff}$.

From these results we conclude that the extremum of the effective
action with respect to $\lambda$
defines a non-trivial saddle point solution and exhibits the condensation of the
Lagrange multiplier at the value $<\lambda> = M$ (this extremum is equivalent
to imposing the constraint (\ref{constrain})~
$g_{ab}=\partial_{a}X_{\mu}\partial_{b}X_{\mu}$ ~). One can also see that the
coupling constant ${1/e^2{}_{eff}}$ is zero at the saddle point and
high derivative term $S_{PK}$ actually is not present.

\section{Perturbation of World-sheet Sigma Model}

One can consider the action (\ref{funaction}) as an example of
nonlinear higher derivative world-sheet sigma model and our aim is to
develop perturbation expansion of this system around some classical
solution and to compute two-loop effective action
\cite{coleman,Iliopoulos:ur,Jackiw:cv,savvidy,batalin,Caswell:gg}.
The action can be represented in the form
\begin{equation}\label{conaction}
S= m\int d^{2}\zeta \sqrt{g}\sqrt{ \left(\Delta(g)
X_{\mu}\right)^{2}},
\end{equation}
here ~$g_{ab}=\partial_{a}X_{\mu}\partial_{b}X_{\mu}$ ~is induced
metric,~$\Delta(g)= 1/\sqrt{g}~\partial_{a}\sqrt{g}g^{ab}
\partial_{b} $ ~is a Laplace operator and
$K^{ia}_{a}K^{ib}_{b}=\left(\Delta(g) X_{\mu}\right)^{2}$. The
second fundamental form $K$ is defined through the relations:
\begin{eqnarray}
K^{i}_{ab}n_{\mu}^{i}&=&\partial_{a}\partial_{b}X_{\mu}-
\Gamma^{c}_{ab}\partial_{c}X_{\mu}=\nabla_{a}\partial_{b}X_{\mu},\quad
\label{extrinc}\\
\quad n_{\mu}^{i}n_{\mu}^{j}&=&\delta_{ij},\quad n_{\mu}^{i}
\partial_{a}X_{\mu}=0,\label{normals}
\end{eqnarray}
where $n_{\mu}^{i}$ are $D-2$ normals and $a,b=1,2; \qquad
\mu=0,1,2,...,D-1; \quad i,j=1,2,...,D-2.$

In order to consider $g$ and $X$
as independent field variables we should introduce standard
Lagrange multipliers $\lambda^{ab}$ and add the corresponding term
to the action \cite{polykov}
\begin{equation}
S_\lambda = -m\int d^2\zeta \lambda^{ab}(\partial_a
X^\mu\partial_b X_\mu -g_{ab})\label{constrain}
\end{equation}
The action $ S + S_\lambda $
is  invariant under two-dimensional general coordinate
transformations and we can fix the conformal gauge
\begin{equation}
g_{ab} = \rho \eta_{ab}
\end{equation}
and should add the corresponding Faddeev-Popov action $S_{FP}$.
The gauge fixed action to be studied is $S+ S_\lambda + S_{\rm FP}.$
We split all fields into a sum of  classical solution plus a quantum
fluctuations
consider:
\begin{eqnarray}
X^\mu &=& X_0 ^\mu + X_1 ^\mu\\
\rho &=& \rho_0 +\rho_1\\
\lambda^{ab} &=& \lambda_0 ^{ab} + \lambda_1 ^{ab}. \label{split}
\end{eqnarray}
In order to investigate the saddle point solution for the Lagrangian
multiplier we shall consider the Ansatz of the form
\begin{equation}
\lambda_0 ^{ab}= \lambda\sqrt{g}g^{ab}=\lambda\eta^{ab}
\end{equation}
where $\lambda$ is a constant field.

Since we are interested in two-loop approximation of the
effective action we have to  expand the action
$S$ up to cubic and quartic interaction in the quantum fields. We
easily get
\begin{eqnarray}
 S + S_\lambda &=& m\int d^2 \zeta \sqrt{\overline{n}^2} +\label{newact} \\
&+& m\int d^2 \zeta \left[ \frac{1}{2\sqrt{\overline{n}^2}}  X_{1}^\mu
\partial^4 (\eta_{\mu\nu} -\frac{\bar{n}_\mu \bar{n}_\nu}
{\bar{n}^2})X_{1}^\nu - \lambda\partial^a X_{1}^\mu \partial_a
X_{1\mu}- 2\lambda_{1}^{ab}\bar{e}_{a}^\mu  \partial_b  X_{1\mu} +
\lambda_{1}\rho_1\right] \nonumber\\
&-& m\int d^2 \zeta \lambda_{1}^{ab}\partial_a X_{1}^\mu
\partial_b X_{1\mu} \nonumber\\
&-& m\int d^2 \zeta \frac{1}{2\sqrt{\bar{n}^2}}\left[(\partial^2 X_1)^2
- \frac{(\partial^2 X_{1}\cdot \bar{n})^2}{\bar{n}^2}\right]\left[
\frac{\bar{n}\cdot\partial^2 X_{1}}{\bar{n}^2} + \frac{(\partial^2
X_1)^2}{4\bar{n}^2} - \frac{5(\partial^2 X_{1}\cdot
\bar{n})^2}{4(\bar{n}^2)^2}\right], \nonumber
\end{eqnarray}
where we have introduced convenient notations:
$\lambda_1 = \eta_{ab}\lambda_1^{ab}$,
$$
\bar{n}^\mu = \partial^2 X_{0}^\mu, \,\,\,\,\, \,\,\,\,\,\,\
\bar{e}^{a}_{\mu}=
\partial^a X_{0\mu},
$$
and used the relation $(\partial_a X_0)^2 = 2\rho_0$. Here we
consider the quantities $\bar{n}^\mu$, $\bar{e}_{a} ^\mu$,
$\rho_0$ to be $\zeta$-independent in contrast to the fast quantities.
Like in \cite{ruben} we shall make an expansion of $X_1^\mu$ fields into
tangential and normal fields $\phi^a$, $\xi^i$:
$$ X_{1\mu} = \phi^a \bar{e}_{{a}_\mu} + \xi^i \bar{n}^{i}_\mu$$
$$\bar{n}^{i\mu} \bar{e}^{a}_\mu = 0\,\,\,\,\,\,\,\,\,\,  \bar{n}^\mu
\bar{e}^{a}_\mu = 0,$$
where the last relation, together with the following ones $
\bar{n}_{\mu}\bar{n}_{\mu}
=g \bar{K}^{ia}_{a}~\bar{K}^{ib}_{b} \equiv g \bar{K}^2,
~\bar{n}_{\mu}\bar{n}^{i}_{\mu}=\sqrt{g}\bar{K}^{ia}_{a}
\equiv \bar{n}^{i},$ can be
easily  seen in conformal gauge.
After the substitution of them into (\ref{newact}) we find that the
relevant terms for our calculation can be collected in the form

\begin{eqnarray} S_1 &=& \frac{m}{2}\int d^2\zeta \left\{\xi^i
\left[\frac{\Pi^{ij}}{\sqrt{\bar{n}^2}}\partial^4 + 2\lambda
\partial^2\delta^{ij}\right]\xi^j +
\phi^a\rho_0\left[\frac{1}{\sqrt{\bar{n}^2}}\partial^4+
2\lambda\partial^2\right] \phi^a
-4\rho_0 \lambda_{1}^{ab}\partial^b \phi^a \right\}\nonumber\\
&+& m\int d^2\zeta \lambda_{1} \rho_1 - m \int d^2 \zeta
\left\{\lambda_{1}^{ab}\partial^a \xi^i
\partial^b \xi^i +\frac{1}{2(\bar{n}^2)^{3/2}}
\left[\bar{n}^i \Pi^{lm}\partial^2\xi^i\partial^2\xi^l\partial^2\xi^m\right]
\right\}\nonumber\\
&-&\frac{m}{8(\bar{n}^2)^{3/2}}\int d^2 \zeta  \Pi^{ij}\partial^2
\xi^i\partial^2\xi^j
\Pi_1^{lm}\partial^2\xi^l\partial^2\xi^m .
\label{newnew}
\end{eqnarray}
All contractions in (\ref{newnew})
have been performed using the tensor $\eta_{ab}$.
Moreover we have denoted by
\begin{equation}
\Pi^{ij} = \delta^{ij}
-\frac{\bar{n}^i\bar{n}^j}{\bar{n}^2}~,~~~~~~~~~~ \Pi_1^{lm}
=\delta^{lm}-5\frac{\bar{n}^l\bar{n}^m}
{\bar{n}^2},~~~~~~~~~~tr \Pi_1 = D-7.
\end{equation}
The operator $\Pi$ is a projector, with the following properties
\begin{equation}
\Pi^2 = \Pi, ~~~~~~~~ \Pi^{ij}\bar{n}^j = 0,~~~~~~~tr \Pi = D-3
\end{equation}
and
\begin{equation}
(\Pi_1 \Pi)^{ij} = (\Pi\,\Pi_1)^{ij} = \Pi^{ij}.
\end{equation}
Let us find now the propagators for all these fields.
The propagators for the $\xi$ fields are easily determined by
inverting the $\xi$$\xi$ part in (\ref{newnew}):

\begin{equation}
<\xi^i(p)\xi^j(-p)> = -i {\sqrt{\bar{n}^2}\over m} \left( {
\delta^{ij} \over p^4-2p^2\lambda\sqrt{\bar{n}^2} } -
\frac{1}{2\lambda\sqrt{\bar{n}^2} (p^2-2\lambda\sqrt{\bar{n}^2})}
\frac{\bar{n}^i \bar{n}^j}{\bar{n}^2}\right)
\end{equation}
where $p$ is momentum. In order to find the propagator for $\phi$,
$\lambda_1$ and $\rho_1$ we use in the momentum space the
decomposition of $\lambda_1$ proposed in \cite{polykov}:
\begin{equation}
\lambda_{1}^{ab}(p) = \omega (p)(\eta^{ab}-\frac{p^a p^b}{p^2}) +
(p^a f^b + p^b f^a -(p\cdot f)\eta^{ab}),~~~~~~tr \lambda_1 =\omega.
\end{equation}
The quadratic part of the action for the fields $\phi$ and
$\lambda_1$ is found to be
\begin{equation}
\int\frac{d^2p}{4\pi^2}
\left\{\frac{m}{2\sqrt{\bar{n}^2}}\rho_0
\phi^a(-p)p^4\phi^a(p) - 2im \rho_0 p^2 \phi^a(-p)f^a(p) -m\lambda
\rho_0 p^2\phi^a(-p)\phi^a (p)\right\},
\end{equation}
consequently we get the following nontrivial propagators:
\begin{equation}
<\phi^a(-p) f^b(p)>= \frac{\eta^{ab}}{2m \rho_0 p^2 }
\end{equation}
\begin{equation}
<f^a(-p)f^b(p)>
=-i\frac{\eta^{ab}}{4m\rho_0}(\frac{1}{\sqrt{\bar{n}^2}}-
\frac{2\lambda}{p^2}).
\end{equation}
It is worth to note here that the propagators for $\phi$ fields
are zero, meaning that longitudinal components of $X_1$
fluctuations are not propagating. For this reason we disregarded
in (\ref{newnew}) all interaction terms depending on $\phi$.
Finally we have to consider the term
\begin{eqnarray*}
m\int\frac{d^2 p}{4\pi^2} \lambda_{1}(-p)\rho_1 (p) &=&
m\int\frac{d^2 p}{4\pi^2}\omega (-p)\rho_1 (p)\\&=& \int\frac{d^2
p}{4\pi^2}[\frac{m}{2}a_{+}(-p)a_{+}(p)
-\frac{m}{2}a_{-}(-p)a_{-}(p)],
\end{eqnarray*}
where we have defined the fields
$$ a_{+} = \omega +\rho_1 \,\,\,\,\,\ a_{-}=\omega-\rho_1. $$
In this way we get the contractions:
\begin{equation}
< a_{+}(-p) a_{+}(p)> = -\frac{i}{m}
\end{equation}
\begin{equation}
<a_{-}(-p) a_{-}(p)> = +\frac{i}{m}.
\end{equation}
In order to regularize divergent integrals we shall use the
dimensional regularization considering our world-sheet action in
$2-2\epsilon$ complex dimensions and shall use the so-called
$\bar{MS}$ scheme in order to renormalize the theory.
All quantities in the above expressions, fields and coupling
constants, are bare quantities. In the two-loop approximation
the effective action shows up divergences  as
poles in $\varepsilon$ up to the order $1/\varepsilon^2$. To
have better control of divergences we shall use explicit
loop expansion of any bare quantity, for example:
\begin{equation}
{\lambda} = \lambda_R + \hbar \delta \lambda_1 +\hbar^2
\delta\lambda_2 ~.
\end{equation}
The dimensionality of the world-sheet fields is as follows:
\begin{eqnarray*}
&&[\xi]=-1\,\,\,\,\,\, [\phi^a]=0\,\,\,\,\,\, [\sqrt{\bar{n}^2}]=-1\\
&&[\rho_1]=-2\,\,\,\,\,\, [\rho_0]=-2 \,\,\,\,\,\ [m]=1\\
&& [\lambda]=1\,\,\,\,\,\,\,\, [\lambda_1]=1~~~~~~~ [p]=0.
\end{eqnarray*}

\section{Tree and one-loop diagrams}
In tree approximation the effective Lagrangian is
\begin{equation}
{{\cal L}}_0 = (m_R + \hbar \delta m_1 + \hbar^2 \delta
m_2)\sqrt{\bar{n}^2}\label{V0}~.
\end{equation}
In the following the subscript indicating the renormalized
quantity will be omitted for brevity of notation.
In the first order, according to \cite{ruben} we have
\begin{eqnarray*}
{{\cal L}}_1 &=& \frac{i}{2}\hbar \ln\det[\Pi^{ij}\partial^4
+ 2(\lambda+\hbar\delta\lambda)\sqrt{\bar{n}^2}\delta^{ij}\partial^2]\\
&=& \frac{i}{2}\hbar (D-3) \int \frac{d^2 k}{(2\pi)^2}\ln(k^2
-2(\lambda+\hbar\delta\lambda)\sqrt{\bar{n}^2}) ~.
\end{eqnarray*}
The integration is regularized in $n=2-2\varepsilon$ complex
dimensions and we get
\begin{equation}
{{\cal L}}_1(\varepsilon) = -\frac{\hbar}{2}\frac{(D-3)\Gamma(\varepsilon-1)}
{(4\pi)^{1-\varepsilon}}
(2\lambda\sqrt{\bar{n}^2})^{1-\varepsilon} .\label{V1}
\end{equation}
 We can expand (\ref{V1}) in $\varepsilon$ and $\hbar$ obtaining
 the following expressions
 \begin{eqnarray}
{{\cal L}}_1(\varepsilon) = -\frac{\hbar}{4\pi}(D-3)\lambda
\sqrt{\bar{n}^2}\left\{\frac{1}{\varepsilon}-\left[\ln \left({\lambda\over
M}\right)-1\right] \right\}  \label{V11}
\end{eqnarray}
together with the counterterm $\delta \lambda$ of order $\hbar^2$
\be
-\frac{\hbar^2}{4\pi}(D-3)\sqrt{\bar{n}^2} ~\delta\lambda ~
\left\{\frac{1}{\varepsilon}-\ln\left({\lambda\over
M}\right)
+\frac{\varepsilon}{2}\left[\ln^2 \left({\lambda\over M}\right)
+{\pi^2 \over 6}\right]\right\}  ,\label{V111}
\ee
where
$$ M= \frac{2\pi e^{-\gamma}}{\sqrt{\bar{n}^2}}$$
and $\gamma$ is the Euler's constant.
The divergence of order $\hbar$ is cancelled by requiring:
\begin{equation}\delta m_1 =
\frac{(D-3)}{4\pi}\frac{\lambda}{\varepsilon}.
\end{equation}
Thus we reproduce the one loop result of \cite{ruben}
\be
{{\cal L}}_1 = \frac{(D-3)}{4\pi}\lambda
\sqrt{\bar{n}^2}\left[ \ln\left({\lambda\over
M}\right)-1\right] .
\ee
The terms of order $\hbar^2$ in (\ref{V0}) and (\ref{V111}) will
provide the counterterms for two-loop diagrams.
In the next section we shall proceed with calculation of
two-loop effective Lagrangian.

\section{Contributions of two-loop diagrams}

Now let us consider two-loop contribution to the effective action.
We are looking for the connected, single-particle irreducible
graphs of order $\hbar^2$ in the expression:
\begin{eqnarray}
&-i&\hbar <0| T\exp\left\{ \frac{i}{\hbar}\int d^2 \zeta
\left[ -\frac{m}{8(\bar{n}^2)^{3/2}}
\Pi^{ij}\partial^2\xi^i\partial^2\xi^j
\Pi_1^{lm}\partial^2\xi^l\partial^2\xi^m  \right. \right.\nonumber\\
&-& \left. \frac{m}{2(\bar{n}^2)^{3/2}} \left[ \bar{n}^i
\Pi^{lm}\partial^2\xi^i\partial^2\xi^l\partial^2\xi^m \right]
\right] \nonumber\\
&+&\left.\frac{i}{\hbar}\int \frac{d^2 p d^2
q}{(2\pi)^4} m \left\{(p^2~q\cdot f(p+q) +
q^2~p\cdot f(p+q))\xi^i (-p)\xi^i (-q)+\right. \right. \nonumber\\
&+&\left. \left. \frac{1}{2}a_+ (p+q)\xi^i (-p)\xi^i
(-q)\left(\frac{(p\cdot q)^2 - p^2 q^2}{(p+q)^2}\right)
+\right.\right.\nonumber\\
&+&\left.\left.\frac{1}{2}a_- (p+q)\xi^i (-p)\xi^i (-q)\left(\frac{(p\cdot q)^2 -
p^2 q^2}{(p+q)^2}\right) \right\} \right\}|0>\label{path}
\end{eqnarray}
Upon rescaling the fields in (\ref{path}) like
$\psi\rightarrow \hbar^{1\over 2}\psi$, expanding the exponential
to the relevant order and applying Wick's theorem, we are left
with three integrals. Let's consider the first one:
\begin{eqnarray*}
\frac{I_1}{\hbar^2} &=&-\frac{3m}{48 (\bar{n}^2)^{3\over
2}}(\Pi^{lm}\Pi_1 ^{ij} +\Pi^{jl}\Pi_1 ^{mi}+\Pi^{li}\Pi_1
^{mj}+\Pi^{mj}\Pi_1
^{li}+\Pi^{im}\Pi_1 ^{lj}+\Pi^{ij}\Pi_1 ^{lm})\\
&& \int\frac{d^2 p
d^2q}{(2\pi)^4}\frac{-i\sqrt{\bar{n}^2}}{m}\left( { \delta^{ij}
\over p^4-2p^2\lambda\sqrt{\bar{n}^2} } -
\frac{1}{2\lambda\sqrt{\bar{n}^2} (p^2-2\lambda\sqrt{\bar{n}^2})}
\frac{\bar{n}^i \bar{n}^j}{\bar{n}^2}\right)p^4 q^4\\
&&\left( { \delta^{lm} \over q^4-2q^2\lambda\sqrt{\bar{n}^2} } -
\frac{1}{2\lambda\sqrt{\bar{n}^2} (q^2-2\lambda\sqrt{\bar{n}^2})}
\frac{\bar{n}^l
\bar{n}^m}{\bar{n}^2}\right)\frac{-i\sqrt{\bar{n}^2}}{m}=
\end{eqnarray*}

\begin{eqnarray*}
&=& \frac{1}{8m\sqrt{\bar{n}^2}}\int\frac{d^2 p d^2
q}{(2\pi)^4}\left[(D-3)\Pi_1 ^{ij} + \frac{2\lambda\sqrt{\bar{n}^2}
(D-3)\Pi_1 ^{ij}}{p^2-2\lambda\sqrt{\bar{n}^2}} \right.\\
&+&\left. (D+1)\Pi^{ij} + \frac{4p^2}{2\lambda\sqrt{\bar{n}^2}}\Pi^{ij} +
\frac{2(D+1)\lambda\sqrt{\bar{n}^2}\Pi^{ij}}{p^2-2\lambda\sqrt{\bar{n}^2}}
\right]\\
&&\left[\delta^{ij} +
\frac{2\lambda\sqrt{\bar{n}^2}}{q^2-2\lambda\sqrt{\bar{n}^2}}\delta^{ij}
-\frac{q^4}{2\lambda\sqrt{\bar{n}^2}(q^2-2\lambda\sqrt{\bar{n}^2})}
\frac{\bar{n}^i \bar{n}^j}{\bar{n}^2}\right].
\end{eqnarray*}
Let us evaluate now this integral in $n=2-2\varepsilon$ complex
dimensions. Using the result that
\begin{equation}
\forall \alpha\geq 0\,\,\,\,\,\,\,\,\, \int d^n p (p^2)^\alpha = 0
\label{prop1}
\end{equation}
we get
\begin{eqnarray*}
\frac{I_1}{\hbar^2} &=& \frac{1}{8m\sqrt{\bar{n}^2}} \int
\frac{d^n p d^nq}{(2\pi)^{2n}}\left[\frac{4\bar{n}^2 \lambda^2
(D-3)(D-7) + 24\bar{n}^2 \lambda^2
(D-3)}{(p^2-2\lambda\sqrt{\bar{n}^2})(q^2-2\lambda\sqrt{\bar{n}^2})}
\right]\\ &=& \frac{\lambda^2 \sqrt{\bar{n}^2}}{2m}(D-1)(D-3)\int
\frac{d^n p d^nq}{(2\pi)^{2n}}
\frac{1}{(p^2-2\lambda\sqrt{\bar{n}^2})(q^2-2\lambda\sqrt{\bar{n}^2})}
\end{eqnarray*}
and due to
\begin{equation}
\left[ \int \frac{d^n
p}{(2\pi)^n}\frac{1}{p^2-2\lambda\sqrt{\bar{n}^2}}\right ]^2 =
-\frac{\Gamma^2(\varepsilon)(4\pi)^\varepsilon
(2\lambda\sqrt{\bar{n}^2})^{-2\varepsilon}}{16 \pi^2}
\end{equation}
we obtain finally:
\begin{equation}
I_1=- \hbar^2\frac{\lambda^2 \sqrt{\bar{n}^2}}{32m \pi^2}(D-1)(D-3)
 \Gamma^2(\varepsilon)(4\pi)^{2\varepsilon}
(2\lambda\sqrt{\bar{n}^2})^{-2\varepsilon}\label{I1}.
\end{equation}
We shell perform the $\varepsilon$ expansion of $I_1$ later.
Now let us consider the second integral
coming from the expansion of (\ref{path}):
\begin{eqnarray*}
\frac{I_2}{\hbar^2} &=& \frac{6}{8(\bar{n}^2)^{3\over
2}m}\left(\frac{\bar{n}^i \Pi^{lm} + \bar{n}^m \Pi^{il}+ \bar{n}^l
\Pi^{mi}}{3}\right)\left(\frac{\bar{n}^j \Pi^{rs} + \bar{n}^r
\Pi^{js}+ \bar{n}^s \Pi^{rj}}{3}\right)\\
& &\int \frac{d^2 p d^2q}{(2\pi)^4}\left[
-\frac{p^2}{2\lambda\sqrt{\bar{n}^2}}\delta^{ij} +
\Pi^{ij}\left(1+
\frac{2\lambda\sqrt{\bar{n}^2}}{p^2-2\lambda\sqrt{\bar{n}^2}}+
\frac{p^2}{2\lambda\sqrt{\bar{n}^2}}\right)\right]\\
& & \left[ -\frac{q^2}{2\lambda\sqrt{\bar{n}^2}}\delta^{lr} +
\Pi^{lr}\left(1+
\frac{2\lambda\sqrt{\bar{n}^2}}{q^2-2\lambda\sqrt{\bar{n}^2}}+
\frac{q^2}{2\lambda\sqrt{\bar{n}^2}}\right)\right]\\
& & \left[ -\frac{(p+q)^2}{2\lambda\sqrt{\bar{n}^2}}\delta^{ms} +
\Pi^{ms}\left(1+
\frac{2\lambda\sqrt{\bar{n}^2}}{(p+q)^2-2\lambda\sqrt{\bar{n}^2}}+
\frac{(p+q)^2}{2\lambda\sqrt{\bar{n}^2}}\right)\right].
\end{eqnarray*}
If we define
\begin{equation}
g(p^2)= 1+
\frac{2\lambda\sqrt{\bar{n}^2}}{p^2-2\lambda\sqrt{\bar{n}^2}}+
\frac{p^2}{2\lambda\sqrt{\bar{n}^2}} ,
\end{equation}
this diagram can be written as:
\begin{eqnarray*}
\frac{I_2}{\hbar^2} &=& -\frac{D-3}{12(\bar{n}^2)^{3\over 2}m}\int
\frac{d^2 p d^2q}{(2\pi)^4}\left[ \frac{3}{8}\frac{p^2 q^2 (p+q)^2
}{\lambda^3 \sqrt{\bar{n}^2}} - \frac{p^2 g(q^2) (p+q)^2
}{2\lambda^2}\right.\\ &-& \left.\frac{q^2 g(p^2) (p+q)^2
}{2\lambda^2} + \frac{g(p^2) g(q^2)(p+q)^2 \bar{n}^2
}{2\lambda \sqrt{\bar{n}^2}}- \frac{p^2 g((p+q)^2)q^2
}{2\lambda^2}\right.\\
&+&\left. \frac{p^2 g(q^2) g((p+q)^2)\sqrt{\bar{n}^2}
}{2\lambda} + \frac{q^2 g(p^2) g((p+q)^2)\sqrt{\bar{n}^2}
}{2\lambda}\right].
\end{eqnarray*}
Using the property (\ref{prop1}) of dimensional
regularization this diagram  amounts to:
\begin{eqnarray*}
\frac{I_2}{\hbar^2} &=& -\frac{(D-3)}{12\sqrt{\bar{n}^2}m}\int
\frac{d^n p d^n q}{(2\pi)^{2n}}\left[\frac{8\bar{n}^2
\lambda^2}{(p^2-2\lambda\sqrt{\bar{n}^2})
(q^2-2\lambda\sqrt{\bar{n}^2})}\right.\\
&+&~~~~~~~~~~~~\left.\frac{p^2}{(p+q)^2-2\lambda\sqrt{\bar{n}^2}} +
\frac{q^2}{(p+q)^2-2\lambda\sqrt{\bar{n}^2}}
\right.\\
&+&\left.\frac{2
\lambda\sqrt{\bar{n}^2}~p^2}{(q^2-2\lambda\sqrt{\bar{n}^2})
((p+q)^2-2\lambda\sqrt{\bar{n}^2})}
+\frac{2 \lambda\sqrt{\bar{n}^2}~q^2}
{(p^2-2\lambda\sqrt{\bar{n}^2})((p+q)^2-2\lambda\sqrt{\bar{n}^2})}
\right].
\end{eqnarray*}
Due to the following general formulas:
\begin{equation}
\int d^n k \frac{1}{(-k^2-2p\cdot k +C)^\alpha}=
i\frac{\pi^{n\over 2}}{\Gamma(\alpha)}(C+p^2)^{{n\over 2}
-\alpha}\Gamma(\alpha-{n\over 2})\label{prop2}
\end{equation}
and
\begin{equation}
\int d^n k \frac{k^a k^b}{(-k^2-2p\cdot k +C)^\alpha}=
i\frac{\pi^{n\over 2}(C+p^2)^{{n\over 2}-\alpha}}{\Gamma(\alpha)}
\left[  \Gamma(\alpha-{n\over 2}) p^a p^b -
\Gamma(\alpha -1 -{n\over 2}){(C+p^2)\over 2} \eta^{ab}
\right]\label{prop3}
\end{equation}
and due to (\ref{prop1}) we get:
\begin{equation}
-\frac{(D-3)}{3m\sqrt{\bar{n}^2}}\int \frac{d^n p d^n
q}{(2\pi)^{2n}}\left[\frac{2\bar{n}^2\lambda^2}
{(p^2-2\lambda\sqrt{\bar{n}^2})(q^2-2\lambda\sqrt{\bar{n}^2})} +
\frac{ p^2\lambda\sqrt{\bar{n}^2}}{(q^2-2\lambda\sqrt{\bar{n}^2})
((p+q)^2-2\lambda\sqrt{\bar{n}^2})}\right].
\end{equation}
Since
\begin{eqnarray*}
&&\int \frac{d^n p d^n q}{(2\pi)^{2n}}\frac{1}{(q^2)^{\nu_3}(p^2-
b^2)^{\nu_1}[(p+q)^2 -b^2]^{\nu_2}} =\\
&&\frac{i^{2+2n}(-b^2)^{n-\nu_1 -\nu_2
-\nu_3}}{\Gamma(\nu_1)\Gamma(\nu_2)\Gamma({n\over
2})(4\pi)^n}\frac{\Gamma(\nu_1 +\nu_2 +\nu_3 -n)\Gamma({n\over
2}-\nu_3)}{\Gamma(\nu_1 +\nu_2 +2\nu_3 -n)}\Gamma(\nu_1 +\nu_3 -
{n\over 2})\Gamma(\nu_2 +\nu_3 -{n\over 2})\label{prop4}
\end{eqnarray*}
and due to (\ref{prop2}) we obtain the following result
\begin{equation}
I_2 =
\hbar^2\frac{(D-3)\sqrt{\bar{n}^2}\lambda^2}{8m\pi^2}\Gamma^2
(\varepsilon)(4\pi)^{2\varepsilon}(2\lambda
\sqrt{\bar{n}^2})^{-2\varepsilon}\label{I2}.
\end{equation}
We can find now the sum of the first two integrals (\ref{I1}) and
(\ref{I2}) of the two-loop effective Lagrangian in the form
\begin{equation}
I_1 + I_2 =- \hbar^2\frac{\lambda^2
\sqrt{\bar{n}^2}(D-3)(D-5)}{32m\pi^2}\Gamma^2
(\varepsilon)(4\pi)^{2\varepsilon}
(2\lambda\sqrt{\bar{n}^2})^{-2\varepsilon} .
\end{equation}

Now let us consider the last non-trivial contribution of order
$\hbar^2$ coming from (\ref{path}). It amounts to the integral
\begin{eqnarray*}
I_3 &=& i\hbar^2 \int \frac{d^2 p
d^2q}{(2\pi)^4}\left[ \frac{-i\sqrt{\bar{n}^2}}{m}
\left(\frac{\delta^{ij}}{p^2}-
\frac{1}{2\lambda\sqrt{\bar{n}^2}}\frac{\bar{n}^i
\bar{n}^j}{\bar{n}^2}\right)
\frac{1}{p^2-2\lambda\sqrt{\bar{n}^2}}\right]~
(-im)\delta^{il}(q_c p^2 + p_c q^2)\\
&&\left[ \frac{-i\sqrt{\bar{n}^2}}{m}\left(\frac{\delta^{lr}}{q^2}-
\frac{1}{2\lambda\sqrt{\bar{n}^2}}\frac{\bar{n}^l
\bar{n}^r}{\bar{n}^2}\right)\frac{1}{q^2-2\lambda\sqrt{\bar{n}^2}} \right]
i m\delta^{jr}(p^2 q_d +q^2 p_d)
\left[\frac{-i\eta^{cd}}{4m\rho_0}\left(\frac{1}{\sqrt{\bar{n}^2}}-
\frac{2\lambda}{p^2}\right)\right]\\
&=&-\hbar^2 \frac{\sqrt{\bar{n}^2}}{4m\rho_0}\int \frac{d^2 p d^2
q}{(2\pi)^4}\left(\frac{D-2}{q^2} + \frac{D-2}{p^2} -
\frac{1}{\lambda\sqrt{\bar{n}^2}} -
\frac{q^2}{2p^2\lambda\sqrt{\bar{n}^2}}
-\frac{p^2}{2q^2\lambda\sqrt{\bar{n}^2}} + \frac{p^2
+q^2}{4\bar{n}^2\lambda^2}\right)\\
&&\left(1 +
\frac{2\lambda\sqrt{\bar{n}^2}}{q^2-2\lambda\sqrt{\bar{n}^2}}
+\frac{2\lambda\sqrt{\bar{n}^2}}{p^2-2\lambda\sqrt{\bar{n}^2}}
+\frac{4\bar{n}^2\lambda^2}{(p^2-2\lambda\sqrt{\bar{n}^2})(
q^2-2\lambda\sqrt{\bar{n}^2})}\right) +\\
&+&\hbar^2 \frac{\bar{n}^2 \lambda}{2m\rho_0}\int \frac{d^2 p d^2
q}{(2\pi)^4}\left(\frac{D-2}{p^2
q^2}-\frac{1}{2p^2\lambda\sqrt{\bar{n}^2}}
-\frac{1}{2q^2\lambda\sqrt{\bar{n}^2}} +
\frac{1}{4\bar{n}^2 \lambda^2}\right)\\
&&\left(1 +
\frac{2\lambda\sqrt{\bar{n}^2}}{q^2-2\lambda\sqrt{\bar{n}^2}}
+\frac{2\lambda\sqrt{\bar{n}^2}}{p^2-2\lambda\sqrt{\bar{n}^2}}
+\frac{4\bar{n}^2\lambda^2}{(p^2-2\lambda\sqrt{\bar{n}^2})
(q^2-2\lambda\sqrt{\bar{n}^2})}\right).
\end{eqnarray*}
Using the property (\ref{prop1}) of  dimensional regularization
and after simple algebraic calculations this graph reduces
to the following integral
\begin{eqnarray}
I_3 &=& -\hbar^2\frac{\lambda\bar{n}^2 (D-3)}{2m\rho_0}\int
\frac{d^n p d^n
q}{(2\pi)^{2n}}\frac{1}{(p^2-2\lambda\sqrt{\bar{n}^2})
(q^2-2\lambda\sqrt{\bar{n}^2})}\nonumber\\
&=&\hbar^2 \frac{\lambda\bar{n}^2(D-3)}{32m\pi^2\rho_0}\Gamma^2
(\varepsilon)(4\pi)^{2\varepsilon}(2\lambda\sqrt{\bar{n}^2})^
{-2\varepsilon}.
\end{eqnarray}
There are also two integrals in  (\ref{path}) of order
$\hbar^2$,  which cancel each other. They are
\begin{eqnarray*}
I_{\pm} &=& \pm\hbar^2\frac{i}{2}\int \frac{d^2 p
d^2q}{(2\pi)^4}\left(\frac{-i\sqrt{\bar{n}^2}}{m}\right)\left( {
\delta^{ij} \over p^4-2p^2\lambda\sqrt{\bar{n}^2} } -
\frac{1}{2\lambda\sqrt{\bar{n}^2} (p^2-2\lambda\sqrt{\bar{n}^2})}
\frac{\bar{n}^i \bar{n}^j}{\bar{n}^2}\right)\\
& & m^2\frac{[(p\cdot q)^2 -p^2
q^2]^2}{(p+q)^4}\delta^{ij}\delta^{lm}\left(\frac{-i\sqrt{\bar{n}^2}}{m}
\right)\left( { \delta^{lm} \over {p^4-2p^2\lambda\sqrt{\bar{n}^2}
}} - \frac{1}{2\lambda\sqrt{\bar{n}^2}
(p^2-2\lambda\sqrt{\bar{n}^2})} \frac{\bar{n}^l
\bar{n}^m}{\bar{n}^2}\right).
\end{eqnarray*}
Therefore the pure two-loop contribution to the effective
Lagrangian is:
\begin{equation}
{{\cal L}}_2 (\varepsilon)=I_1 +I_2 +I_3 = -\hbar^2\left[\frac{\lambda^2
\sqrt{\bar{n}^2}(D-3)(D-5)}{32m\pi^2}-\frac{\lambda\bar{n}^2(D-3)}{32m\pi^2\rho_0}
\right] \Gamma^2 (\varepsilon)(4\pi)^{2\varepsilon}
(2\lambda\sqrt{\bar{n}^2})^{-2\varepsilon}.
\end{equation}
The Laurent expansion around $\varepsilon=0$ is
\begin{equation}
{{\cal L}}_2 (\varepsilon)=  -\hbar^2
\frac{(D-3)}{32\pi^2 } {\lambda \over m}\sqrt{\bar{n}^2}
\left[(D-5)\lambda
-\frac{\sqrt{\bar{n}^2}}{\rho_0}\right]
\left\{\frac{1}{\varepsilon^2}-
\frac{2}{\varepsilon}\ln\left(\frac{\lambda}{M}\right)
+2\ln^2\left(\frac{\lambda}{M}\right)
+\frac{\pi^2}{6}\right\}.\label{two}
\end{equation}
Summing all contributions in $\hbar^2$-order coming from
(\ref{V11}) and (\ref{two}) we can cancel the non-polynomial
divergencies first by requiring
\begin{equation}
\hbar^2 \frac{(D-3)}{4\pi} \delta\lambda ~\sqrt{\bar{n}^2}
\ln\left(\frac{\lambda}{M}\right) +  \hbar^2
\frac{(D-3)[(D-5)\lambda
-\frac{\sqrt{\bar{n}^2}}{\rho_0}]}
{16\pi^2 \varepsilon} {\lambda\over m}\sqrt{\bar{n}^2}
\ln\left(\frac{\lambda}{M}\right)=0 ,
\end{equation}
that is
\be
\delta\lambda= -  {1 \over  4 \pi \varepsilon}~ \frac{\lambda}{ m }
~\left[(D-5)\lambda -\frac{\sqrt{\bar{n}^2}} {\rho_0}\right] \label{count1}
\ee
and then the polynomial divergencies requiring that
\begin{equation}
\delta m_2 =  -{1\over \varepsilon^2}~
\frac{(D-3)}{32\pi^2}{\lambda\over m} ~\left[(D-5)\lambda
-\frac{\sqrt{\bar{n}^2}}{\rho_0}\right].
\label{count2}
\end{equation}
Finally substituting into ${{\cal L}}_{eff}={{\cal L}}_0(\varepsilon) +
{{\cal L}}_1(\varepsilon)+ {{\cal L}}_2 (\varepsilon)$
the values found for $\delta\lambda$ and $\delta m_2$ from the
equations (\ref{count1}) and (\ref{count2}) we get the
finite $\bar{MS}$ world-sheet effective
Lagrangian in two-loop order:
\beqa
{{\cal L}}_{2} = -
\frac{(D-3)(D-5)}{32\pi^2}
~{\lambda^2 \over m}\sqrt{\bar{n}^2}~
\ln^2 \left(\frac{\lambda}{M}\right) \label{resul}
\eeqa
together with the term which we shall write separately, because,
as it is easy to see, it is proportional to the Polyakov-Kleinert
Lagrangian
\be
{{\cal L}}_{2~PK} =
\frac{(D-3) }{16\pi^2 }{\lambda\over m}   ~\ln^2
\left(\frac{\lambda}{M}\right) {\bar{n}^2 \over \rho_0}~~
\propto ~~{(\partial^{2} X^{\mu})^{2} \over \rho }
=\sqrt{g}K^{ia}_{a}K^{ib}_{b} . \label{resul1}
\ee
Quantum fluctuations generate high powers of
the extrinsic curvature tensor $K^{i}_{ab}$ in the form of
Polyakov-Kleinert action with the induced coupling constant
\be
{1\over e^{2}_{eff} }=\frac{(D-3) }{16\pi^2 }{\lambda \over m}  ~\ln^2
\left(\frac{\lambda}{M}\right) . \label{resul2}
\ee
From these results we conclude that the extremum of the effective
action with respect to $\lambda$
defines a non-trivial saddle point solution and exhibits the condensation of the
Lagrange multiplier at the value $<\lambda> = M$ (this extremum is equivalent
to imposing the constraint (\ref{constrain})
$g_{ab}=\partial_{a}X_{\mu}\partial_{b}X_{\mu}$). One can also see that the
coupling constant ${1/e^2{}_{eff}}$ is zero at the saddle point and
high derivative term ${{\cal L}}_{PK} $ actually is not present. The absence
of high derivative terms in the effective action can also be seen
in physical gauge calculations \cite{ruben}, there are two terms
(see formulas (50) and (52) in \cite{ruben}) proportional
to ${{\cal L}}_{PK} $, but they cancel each other.

One of the authors (R.F.) is indebted to the Demokritos
National Research Center for support and kind hospitality. This
work was supported in part by the by EEC Grant HPRN-CT-1999-00161.

\vfill
\end{document}